\documentclass[journal]{IEEEtran}
\pdfoutput=1

\usepackage{amsmath,amsfonts}
\usepackage{multirow}
\usepackage{graphicx}
\usepackage{textcomp}
\usepackage{xcolor}
\usepackage{framed}
\usepackage{listings}
\usepackage{graphicx,subfigure}
\usepackage{bbding}
\usepackage{color}
\usepackage{makecell}
\usepackage{enumitem}
\usepackage{xspace}
\usepackage{threeparttable}
\usepackage{verbatim}
\usepackage{hyperref}
\usepackage{pifont}
\usepackage{cleveref}
\usepackage{colortbl}
\usepackage{booktabs}
\usepackage{amsmath}
\usepackage{xspace}
\usepackage{epstopdf}
\usepackage[vlined, ruled,linesnumbered]{algorithm2e}
\usepackage[font=small,skip=2pt]{caption}

\def\eg{\emph{e.g.}} 
\def\ie{\emph{i.e.}}

\def\etal{\emph{et al.}}

\newtheorem{definition}{Definition}

\newcommand{\red}[1]{\textcolor{black}{#1}\xspace}
\newcommand{\rred}[1]{\textcolor{red}{#1}\xspace}

\DeclareMathOperator*{\argmax}{arg\,max}
\DeclareMathOperator*{\argmin}{arg\,min}

\newcommand{\tool}{\textit{TokenProber}\xspace}

\newcommand{\distance}{2pt}
\AtBeginDocument{%
  }

\begin{document}



\title{\tool: Jailbreaking Text-to-image Models via Fine-grained Word Impact Analysis}

\author{Longtian Wang,~\IEEEmembership{Graduate Student Member,~IEEE,}
\thanks{Longtian Wang, Yuhan Zhi, and Chao Shen are with the School of Cyber Science and Engineering, Xi'an Jiaotong University, Xi'an 710049, China (e-mail: chaoshen@mail.xjtu.edu.cn).}
        Xiaofei Xie,~\IEEEmembership{Member,~IEEE,}
        \thanks{Xiaofei Xie is with Singapore Management University, Singapore 188065 (e-mail: xfxie@smu.edu.sg).}
        Tianlin Li,~\IEEEmembership{Member,~IEEE,}
        \thanks{Tianlin Li is with the Nanyang Technological University, Singapore 639798.}
        Yuhan Zhi,~\IEEEmembership{Graduate Student Member,~IEEE,}
        and~Chao Shen,~\IEEEmembership{Senior Member,~IEEE}
        
        }




\maketitle

\begin{abstract}
Text-to-image (T2I) models have significantly advanced in producing high-quality images. However, {such models have the ability to generate images containing not-safe-for-work (NSFW) content, such as pornography, violence, political content, and discrimination.
To mitigate the risk of generating NSFW content, refusal mechanisms, \ie, safety checkers, have been developed to check potential NSFW content.} Adversarial prompting techniques have been developed to evaluate the robustness of the {refusal mechanisms}. 
The key challenge remains to subtly modify the prompt in a way that preserves its sensitive nature while bypassing the refusal mechanisms.
In this paper, we introduce \tool, a method designed for sensitivity-aware differential testing, {aimed at evaluating the robustness of the refusal mechanisms in T2I models by generating adversarial prompts.}
Our approach is based on the key observation that adversarial prompts often succeed by exploiting discrepancies in how T2I models and safety checkers interpret sensitive content. Thus, we conduct a fine-grained analysis of the impact of specific words within prompts, distinguishing between \textit{dirty} words that are essential for NSFW content generation and \textit{discrepant} words that highlight the different sensitivity assessments between T2I models and safety checkers. Through the sensitivity-aware mutation, \tool generates adversarial prompts, striking a balance between maintaining NSFW content generation and evading detection. Our evaluation of \tool against 5 safety checkers {on 3 popular T2I models}, using 324 NSFW prompts, demonstrates its superior effectiveness in bypassing safety filters compared to existing methods (\eg, 54\%+ increase on average), 
{highlighting \tool's ability to uncover robustness issues in the existing refusal mechanisms.} 
The source code, datasets, and experimental results are available in~\cite{noauthor_anonymized_nodate}. 

\rred{\textbf{Warning:} This paper contains model outputs that are offensive in nature.}
\end{abstract}


\IEEEpeerreviewmaketitle

\section{Introduction} \label{introduction}

The Text-to-Image (T2I) models have gained widespread attention due to their excellent capability in synthesizing high-quality images. 
T2I models, such as Stable Diffusion~\cite{rombach2022high} and DALL·E~\cite{ramesh2022hierarchical}, process the textual descriptions provided by users, namely prompts, and output images that match the descriptions. 
Such models have been widely used to generate various types of images, 
for example, 
the Lexica~\cite{noauthor_lexica_nodate} contains more than five million images generated by Stable Diffusion. 
Besides, there are many advanced applications based on T2I models, for instance, Imagic~\cite{kawar2023imagic} and Sine~\cite{zhang2023sine} allow users to edit objects in the source image using text descriptions, Textual inversion~\cite{gal2022image} and DreamBooth~\cite{ruiz2023dreambooth} fine-tune T2I models to generate images of special unique concepts.

With the popularity of T2I models, ethical concerns about the safety of synthesized image content have gradually emerged, \ie, T2I models may synthesize images containing not-safe-for-work (NSFW) content, {such as pornography, violence, political content, and discrimination}. To prevent T2I models from synthesizing offensive or inappropriate images, {model developers often add refusal mechanisms to the model to refuse the generation of such NSFW content}. The typical approach to achieve the refusal mechanisms is to deploy safety checkers to block the NSFW prompts and corresponding generations. However, it has been reported that these safety checkers can be bypassed by manually crafted malicious prompts \cite{washingtonpost2023aiabuse,guardian2024aiabuse}, highlighting the need for a thorough robustness evaluation of the safety checkers in T2I models.

\begin{figure}[!t]
    \centering
    \includegraphics[width=\linewidth]{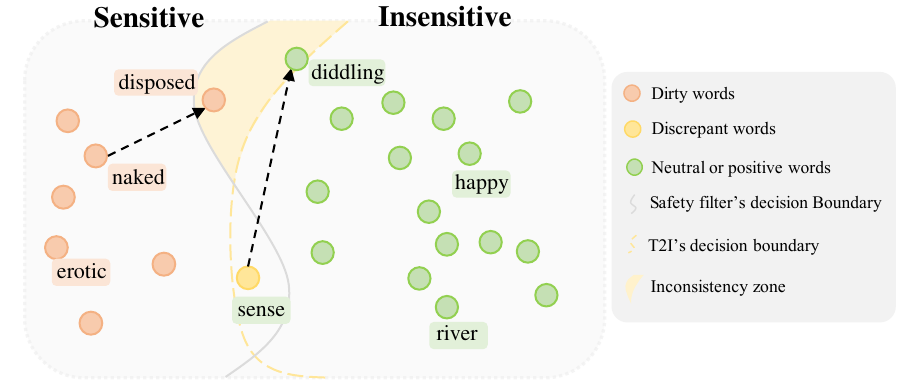}
    \caption{Influence of different words within prompts. {The explicit use of dirty words (\eg, `naked') can lead T2I models to generate NSFW content, which simultaneously triggers detection by the safety checkers. However, due to differing decision boundaries between the T2I model and the safety checkers regarding NSFW content, some words (\eg, `diddling') may fall within the region where discrepancies exist, specifically in the overlap between the boundaries of the safety checker and the T2I model. These words can cause the T2I model to generate NSFW content without being detected by the safety checker.}}
    \label{fig:illustration}
\end{figure}

{Several adversarial testing methodologies~\cite{qu2023unsafe, rando2022red, yang2024sneakyprompt, ba2023surrogateprompt} have been designed to reveal the highlighted robustness issues. 
These studies aim to reduce the likelihood that prompts containing sensitive words, as well as the resulting generated images, are detected by these safety filters.
To achieve this purpose, the foundational strategy in these previous methods is to substitute sensitive words with less detectable alternatives.
For example, SneakyPrompt~\cite{yang2024sneakyprompt} utilizes reinforcement learning (RL) to explore alternative tokens and formulate new prompts, and SurrogatePrompt~\cite{ba2023surrogateprompt}, using Large Language Models (\eg, ChatGPT) to find alternative words for NSFW parts in prompts to bypass Midjourney's safety checkers.
However, we find that such strategies remain ineffective because they tend to generate non-adversarial prompts (\ie, false positives), where the prompts fail to include NSFW content in the generated images although they manage to evade safety checkers, {\eg, 66\% of prompts reported success by SneakyPrompt are flase positives.} 
The ineffectiveness mainly arises from \textit{their emphasis on modifying explicitly sensitive words in the prompts, which may inadvertently remove the key terms that trigger NSFW content during generation, particularly when trying to bypass detection by refusal mechanisms. Consequently, these previous approaches face a dilemma: removing sensitive words decreases the chances of detection but also diminishes the potential to generate images with NSFW content.}
}

As illustrated in Fig.~\ref{fig:illustration}, the goal of these previous methods can be understood as crafting prompts that fall within the insensitive side of the safety filter's decision boundary, potentially losing their NSFW nature.
To preserve the NSFW nature of the generated content, we refine the goal to crafting adversarial prompts that exploit the discrepancy between the safety checker's and the T2I model's decisions (as highlighted in yellow in Fig.~\ref{fig:illustration}): the safety checker fails to classify the prompt and the generation as unsafe, yet the T2I model proceeds to produce sensitive images with NSFW content. 
\red{This discrepancy is intuitive given the nature of machine learning. Two distinct models are expected to yield differing outputs for certain inputs due to inherent differences in their training datasets, architectures, and parameter weights. Unless the models are identical in every aspect, such differences are inevitable.}
However, it is challenging to effectively craft adversarial prompts that exploit this discrepancy.

In our study, we propose \tool, a novel framework designed to assess the robustness of safety checkers within T2I models. To tackle the key challenge outlined above, \tool performs a fine-grained analysis of the influence of various words, enabling us to balance between maintaining NSFW content in the image generated by the T2I model and bypassing the detection of safety checkers (\ie, the prompts falling within the inconsistency zone depicted in Fig.~\ref{fig:illustration}). \tool identifies two key types of words to navigate this balance: \textit{Dirty Words}, which are essential for retaining the NSFW content and should maintain their semantic, and \textit{Discrepant Words}, which are not inherently dirty (\eg, neutral or positive) but significantly affect the safety checker's predictions in a negative manner. 
Discrepant words essentially indicate that mutating these words is more likely to mislead the safety checker.
By mitigating the influence of discrepant words, the checker is more likely to assess the prompt positively. By safeguarding dirty words while diminishing the influence of discrepant words, \tool achieves an optimal equilibrium, which enables the creation of adversarial prompts that exploit the discrepancies between the decision of the safety checker and the T2I model.


Technically, \tool mainly includes two phases: Word-Level Sensitivity Analysis and Sensitivity-Aware Differential Testing. The initial phase analyzes the given prompt at the word level to identify \textit{dirty} words, which are essential for retaining NSFW content, and \textit{discrepant} words, which have a high potential to impact the safety checker's prediction towards negative. 
Following the analysis, the second phase performs a fuzzing, which iteratively mutates both the \textit{dirty} and \textit{discrepant} words towards the decision toward the region of discrepancies between the decision boundaries of the T2I model and the safety checker (as shown in Fig. \ref{fig:illustration}). 
In each iteration, the goal is to alter the \textit{dirty} words of the seed prompts in a way that their inherent NSFW semantic is still preserved, while the \textit{discrepant} words are changed to ones that significantly differ in similarity. 
To exploit the boundary discrepancies more effectively, \tool employs the principles of differential testing. Inspired by \cite{xiao2023latent}, which approximates a surrogate decision boundary for generative models, we build a \red{surrogate safety checker} to classify images as safe or unsafe trained on the generations of T2I models \cite{qu2023unsafe} and use the \red{surrogate safety checker's} decision boundary to approximate that of the T2I models.
We then compare the safety score from the \red{surrogate safety checker} with that of the target safety checker to indicate discrepancies between the decision boundaries. 
Thus, in each iteration, we select prompts maximizing the difference in safety scores between the target checker and the \red{surrogate safety checker} as seed prompts for the next iterations. 
We conduct a comprehensive evaluation to demonstrate the effectiveness of \tool
in generating adversarial prompts. 
{Specifically, we evaluate 5 target safety checkers applying to 3 popular T2I models} and 324 NSFW prompts to evaluate \tool. The results demonstrate 
that 1) \tool has superior effectiveness in bypassing target safety checkers compared to state-of-the-art methods (\eg, 54\%+ increase on average). 2) \tool is more efficient in generating adversarial prompts, \eg, 
on average it takes 60\%+ less time to generate an adversarial prompt. 

To summarize, this paper makes the following contributions:
\begin{itemize}[leftmargin=*]
\item 
We provide an understanding of the impact of two types of words within prompts: \textit{dirty} words, which are integral to generating NSFW content, and \textit{discrepant} words, which are sensitive in impacting the robustness of safety checkers. 
\item 
We introduce a novel {robustness evaluation}
approach \tool that employs word sensitivity-aware differential testing to generate adversarial prompts capable of producing NSFW images. 
\item We conduct extensive experiments to evaluate the effectiveness of \tool on various open-source safety checkers {on various T2I models}. The evaluation results show that \tool achieves a 54\%+ average increase in the bypass rate compared to the state-of-the-art methods.  
\end{itemize}







\section{Preliminary}
\subsection{Text-to-image Model}
\begin{figure*}[!t]
    \centering
    \includegraphics[width=0.9\linewidth]
    {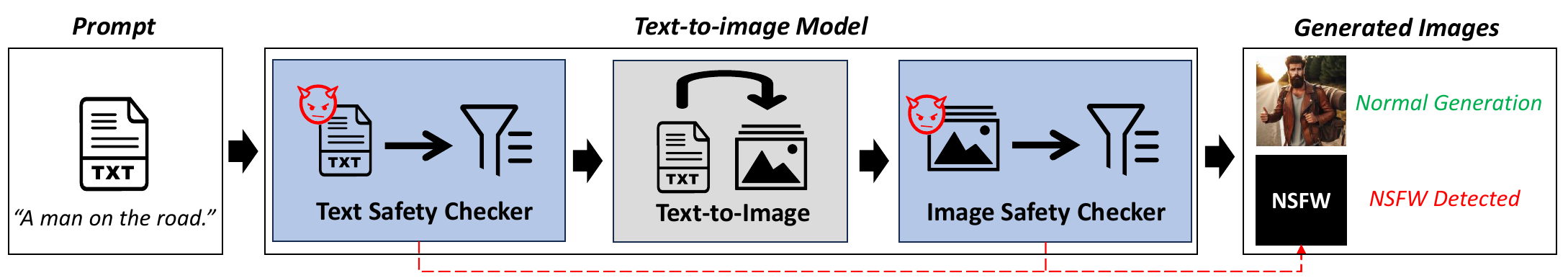}
    \caption{Text-to-image model structure}
    \label{fig:structure}
\end{figure*}

Text-to-image (T2I) models synthesize images conditioned by textual descriptions provided by users (\ie, prompts). Fig.~\ref{fig:structure} shows the structure of a T2I model. Typically, T2I models comprise a language model and an image generation model. The language model, \eg, CLIP's text encoder~\cite{radford2021learning}, serves to comprehend the prompt and convert it into text embeddings to guide image generation. The image generation model then employs a diffusion process, initiating with random noise and progressively denoising conditioned by the text embeddings to synthesize images that align with the user's description.

\subsection{Safety checker for T2I Model}
T2I models have impressive performance in synthesizing high-quality images. Meanwhile, ethical concerns about the safety of the generated image content have gradually emerged. To prevent T2I models from generating not-safe-for-work (NSFW) content, various safety checkers have been proposed. The safety checker, denoted as $SC$, takes a prompt $p$ and the synthesized image $I$ as input, then outputs the probability of containing NSFW content. We called the probability Safety Score, denoted as $SC(p, I)$.

As shown in Fig.~\ref{fig:structure}, the safety checkers can be divided into three categories~\cite{yang2024sneakyprompt}: text safety checkers, image safety checkers, and text-image safety checkers. 

\begin{itemize}[leftmargin=*]
 \item 
\textbf{Text safety checker}: The input of safety checkers in this category is the prompt or the text embedding obtained by language model. Prompt containing sensitive words or embedding with high similarity to predefined unsafe concepts will be blocked before synthesizing images.
\item
\textbf{Image safety checker}: Safety checks in this category takes the synthesized image as input and detect NSFW content. Typically, a binary classifier trained on NSFW images and safe images is employed to detect whether synthesized images contain NSFW content.
\item
\textbf{Text-image safety checker}: Safety checks in this category utilize text embedding of unsafe concept and synthesized images to prevent the generation of NSFW content. To the best of our knowledge, currently only the open-source Stable Diffusion~\cite{rombach2022high} utilizes this type of safety checker. The safety checker firstly employs CLIP image-encoder to obtain the embedding of synthesized image, and blocks image embeddings that have high cosine similarity to 17 pre-defined text embeddings of unsafe concepts~\cite{rando2022red}.
\end{itemize}

\subsection{Problem Definition}\label{sec:problemdef}

\begin{definition} [Robustness of Safety Checker for T2I Model] \label{def-mr}
Given an T2I Model $M$ equipped with a safety checker $F$, for any prompt $p$ that is filtered by the safety checker (\ie, $F(M, p)=1$), indicating that the prompt is recognized as invalid, we define the robustness of $F$ on $p$ as:
\begin{equation}\label{eq:MR}
\forall p' ||M(p)-M(p')||< \epsilon \Longrightarrow F(M, p) = F(M, p').
\end{equation}
\end{definition}
This equation posits that for any variant of the prompt $p'$, if the semantic difference in the outputs of $M$ is small (\ie, $||M(p)-M(p')||< \epsilon$), then the safety checker $F'$s decision should remain consistent for both $p$ and $p'$.

The objective of our testing is to search the adversarial prompt $p'$ that satisfies the following two properties:
\begin{itemize}[leftmargin=*]
 \item
 \textit{P1: NSFW maintenance} (\ie, $||M(p)-M(p')||< \epsilon$): The images synthesized by the generated adversarial prompt should still contain NSFW content. If a prompt that bypasses the safety checker but no longer contains NSFW content, it is not a valid adversarial prompt.
 \item 
 \textit{P2: Bypassing the filter} (\ie, $F(M, p) \neq F(M, p')$: The target safety checker fails to block the prompt or the NSFW content after some perturbations.
\end{itemize}

\section{Methodology}
\subsection{Overview of our Approach}

\begin{figure*}[!t]
    \centering
    \includegraphics[width=0.9\linewidth]{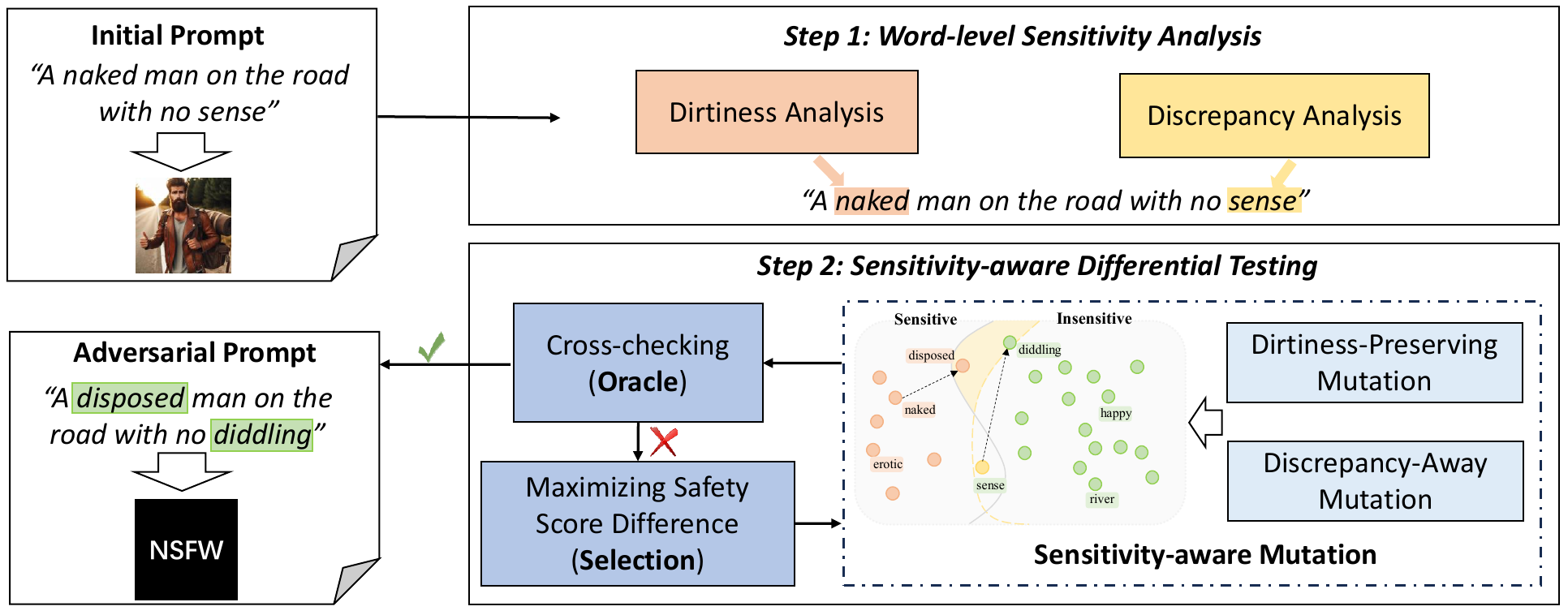}
    \caption{The workflow of \tool} 
    \label{fig:overview}
\end{figure*}

Fig.~\ref{fig:overview} depicts the overview of \tool. \tool takes an initial prompt that describes an NSFW image scene, the T2I model and the target safety checker as inputs. For simplicity of presentation, the model and target safety checker are omitted from the figure. Note that the output of the initial prompt is blocked by the target safety checker under test. \tool operates in two main steps. In the first step, \tool analyzes the impact of individual words on the generation of NSFW content and their influence on the target safety checker's prediction. This involves examining each word for its ``Dirtiness'' and ``Discrepancy''. Dirtiness Analysis assesses whether a word predominantly contributes to NSFW content generation (\eg ``naked''), while Discrepancy Analysis evaluates the word's effect on the target safety checker's robustness.

Based on the insights from the sensitivity analysis, \tool performs a fuzzing strategy to optimize the target prompt. Two mutation strategies are proposed to refine the target prompt. Dirtiness-Preserving Mutation looks for semantically similar words to replace dirty ones, maintaining NSFW content generation potential while possibly challenging the target safety checker's robustness. Discrepancy-Away Mutation aims to replace Discrepant words with alternatives that reduce the likelihood of being classified as unsafe by the target safety checkers, thus improving the prompt's potential to bypass the target safety checkers.

Since no perfect oracle exists to definitively judge whether the image generated from a new prompt contains NSFW content. \tool performs cross-checking with a \red{surrogate safety checker}, which is trained on the outputs of T2I models to approximate a surrogate decision boundary for generative models. A prompt deemed safe by the target safety checker but flagged by the \red{surrogate safety checker} is considered a potential adversarial prompt. The selection of the best candidate prompt from the generated mutations is guided by the objective to maximize the safety score difference between the target safety checkers and \red{surrogate safety checker}. This iterative process continues until a successful adversarial prompt is generated or the maximum number of iterations is reached.

\subsection{Word-level Sensitivity Analysis}
\label{sec:approximation}
In the generation process with T2I models, replacing different words in the prompt leads to varied effects on the NSFW content of the synthesized images and the robustness of the target safety checker. As outlined in Section~\ref{sec:problemdef}, our dual objectives in crafting adversarial prompts are: preserving NSFW content (P1) and evading the safety filter (P2). Identifying words pivotal for both preserving NSFW content and deceiving the target safety checker is essential for achieving these goals. To this end, we perform a fine-grained analysis to evaluate two critical attributes of the word: \textit{dirtiness} and \textit{discrepancy}.

\subsubsection{Dirtiness Analysis}
\textit{Dirtiness Analysis} aims to identify words in the prompt that significantly influence the NSFW content of synthesized images. These words are typically easily identified as sensitive, \eg, ``naked'', ``nude'' and ``sexual'', previous work~\cite{qu2023unsafe} has demonstrated that prompts include such words can retrieve images containing NSFW content. We adopt a simple strategy to identify dirty words based on the existing sensitive word list, which has been used in previous work~\cite{yang2024sneakyprompt}. The sensitive word list includes various themes, such as violence, discrimination, and sexual content, making it suitable for detecting multiple categories of unsafe content. 

Formally, given a prompt $p=\{w_1, w_2,...,w_n\}$ and the predefined NSFW word list $W_{\text{NSFW}}$, the dirtiness of a word $w\in p$ is determined as:
$$Dt(w)=
\left\{
\begin{array}{ll}
1,  & w\in W_{\text{NSFW}} \\
0, &  \mathrm{otherwise.} 
\end{array}
\right.
$$

\subsubsection{Discrepancy Analysis}
While our primary aim is to diminish the negative impact of prompts to meet objective P2, we cannot simply eliminate obvious dirty words required for maintaining objective P1. Instead, our strategy involves identifying the words that, although not dirty words, significantly contribute to the safety checker's assessment, which are called discrepant words. Specifically, discrepant words are highlighted within the discrepant zone of Fig.~\ref{fig:illustration}, where prompts containing these words are interpreted as insensitive by the T2I model yet flagged as sensitive by the target safety checker.

By minimizing the influence of discrepant words, we aim to achieve objective P2. The presence of discrepant words underscores the quality issue in the target safety checker's robustness, revealing that its predictions are unduly affected by words that should not skew its evaluations. Our methodology exploits this flaw, targeting the removal of discrepant words to refine the prompt and improve its acceptance by the target safety checker.




As depicted in Fig.~\ref{fig:illustration}, to quantify the discrepancy of a word, we need to evaluate the difference in the decision between the target safety checker (which may be interpreted as insensitive) and the T2I model (which might be interpreted as sensitive). Given that the T2I model does not offer a direct measure of sensitivity, we opt for a 
\red{surrogate safety checker} trained on the outputs of T2I models to approximate the decision boundary of T2I models. 
 Since the training data for the \red{surrogate safety checker} is entirely composed of images generated by the T2I model, it will better approximate the decision boundary of the T2I model.
 Thus, utilizing a \red{surrogate safety checker} as a reference (denoted as $SC_r$) provides a meaningful comparative basis to identify the discrepant words of the target checker  (denoted as $SC_t$).

 Given a prompt $p=\{w_1, w_2,...,w_n\}$, we use $p\backslash w$ to denote the prompt after removing the word $w$ and $I$ to denote the image synthesized by $p\backslash w$. The discrepancy of a word $w \in p$ is calculated as the safety score difference (\ie, the the probability that the modified prompt $p\backslash w$ predicted by the safety checkers $SC_t$ and $SC_r$), formally expressed as:
$$Dc(w) =SC_{r}(P\backslash w, I) - SC_{t}(P\backslash w, I).$$

Intuitively, if a word $w$ exhibits high discrepancy, it implies that, when $w$ is removed, the modified prompt $p\backslash w$ is deemed less risky by the target safety checker, compared to the \red{surrogate safety checker}. Such discrepancy suggests that reducing the influence of $w$ is particularly crucial for bypassing the target safety checker's scrutiny.




\subsection{Sensitivity-aware Differential Testing}
\label{sec: diff-testing}

\begin{algorithm}[t]
\footnotesize
\caption{\tool}\label{alg:mainalgo}
\SetKwInOut{Const}{Const}
\SetKwInOut{Input}{Input}
\SetKwInOut{Output}{Output}

	\SetKwInOut{Continue}{continue}
	\SetKwProg{myproc}{Procedure}{}{}
\Input{$p$: seed prompt, $SC_t$: target safety checker, $SC_r$: \red{surrogate safety checker}, $DirLis$: the dirty word substitution list, $DisLis$: the discrepant word substitution list.}
\Output{$p'$: potential adversarial prompt}
\Const{$T$: testing budget, $K$: number of discrepant words.}

\BlankLine
$p_{cur} \leftarrow p$\;\label{line:initial1}
$W_{dir} \leftarrow \{w|w\in p \wedge Dt(w)=1\}$\; \label{line:getdirtyword}
$W_{dis} \leftarrow TopK_{Dc}(K, p)$\; \label{line:getdisword}
\For{$t\in [0, T)$}{ \label{line:startfuzzing}
$W_{dir}' \leftarrow RandSelect(W_{dir})$\; \label{line:randselectdir}
$Sub_{dir} \leftarrow  \{(w, \argmax_{w'\in DirLis} Similarity(w, w') )| w\in W_{dir}'\}$ \; \label{line:selectbestdirty}
$p_{dir} \leftarrow Replace(p_{cur}, Sub_{dir})$\;  \label{line:dirreplace}

$W_{dis}' \leftarrow RandSelect(W_{dis})$\; \label{line:randselectdis}
$Sub_{dis} \leftarrow  \{(w, \argmin_{w'\in DisLis} Similarity(w, w')) | w\in W_{dis}'\}$ \; \label{line:selectbestdis}
$p_{dis} \leftarrow Replace(p_{dir}, Sub_{dis})$\;  \label{line:disreplace}

$BestFit \leftarrow fitness(p_{cur})$ \;  \label{line:initfit}
\For{$p'\in \{p_{dis}, p_{dir}\}$}{
    \If{$InValid(p')$}{ \label{line:isvalid}
        \textbf{return} $p'$ \;\label{line:returnresult}
    }
    \If{$fitness(p')>BestFit$}{\label{line:compare}
        $BestFit \leftarrow fitness(p')$ \;  \label{line:updatefit}
        $p_{cur} \leftarrow p'$ \;  \label{line:updatecurprompt}
    }
}
}

\textbf{return} $p_{cur}$ \;\label{line:failresult}
\end{algorithm}

Building on the foundation of word-level sensitivity analysis, \tool employs a greedy-based fuzzing strategy to identify potential adversarial prompts. This approach involves mutation of specific word categories through \textit{Dirtiness-Preserving Mutation} and \textit{Discrepancy-Away Mutation}. As detailed in Algorithm~\ref{alg:mainalgo}, \tool operates with an array of inputs: an initial prompt $p$, a target safety checker $SC_t$, a \red{surrogate safety checker} $SC_r$, and lists of candidate substitutions for dirty ($DirLis$) and discrepant ($DisLis$) words, aiming to generate the adversarial prompt $p'$. Note that the initial prompt can be blocked by the target solver.
There are some hyperparameters that can be adjusted, such as the testing budget $T$, indicating the maximum number of iterations, and the focus on the top $K$ discrepant words in the prompt.

$p_{cur}$ represents the current prompt in each iteration, which is initialized as $p$ (Line~\ref{line:initial1}). The process begins by identifying all dirty words within the prompt $p$ (Line~\ref{line:getdirtyword}), along with the top 
$K$ discrepant words based on their discrepancy scores (Line~\ref{line:getdisword}). These discrepant words are then ordered by their impact on the safety checker's prediction, prioritizing those with higher discrepancy scores for further process. With these selections in place, \tool proceeds to conduct a greedy search operation within the specified budget (Line~\ref{line:startfuzzing}), aiming to refine $p_{cur}$ by adjusting the presence and influence of these critical word types. 

\subsubsection{Dirtiness-Preserving Mutation}
During each iteration, \tool firstly mutates the dirty words, adhering to the strategy of maintaining the negative semantics integral for NSFW content generation. The fundamental idea is to substitute a dirty word with another that holds similar negative connotations, thereby preserving the prompt's NSFW nature while potentially challenging the safety checker's robustness.

\tool randomly selects a subset of dirty words from the current prompt $p_{cur}$ for mutation (Line~\ref{line:randselectdir}). For each selected dirty word, \tool evaluates the semantic similarity with candidates from $DirLis$, aiming to find the most closely aligned substitute in terms of dirtiness (Line~\ref{line:selectbestdirty}). This similarity assessment relies on the cosine similarity between their word embeddings obtained by a language model, calculated as $Cosine\_Similarity(embed(w), embed(w'))$. 
The dirty words in $p_{cur}$ are then replaced with their most semantically similar counterparts from $DirLis$, resulting in a new prompt $p_{dir}$ (Line~\ref{line:dirreplace}).

\subsubsection{Discrepancy-Away Mutation}
Subsequently, \tool targets discrepant words for mutation, aiming to select substitutions that deviate maximally from the original, thereby minimizing the discrepant word's impact and potentially creating a prompt that is more likely to bypass the safety filter.
Similarly, \tool randomly chooses a group of discrepant words within the current prompt for the mutation (Line~\ref{line:randselectdis}). For every discrepant word in this subset, \tool evaluates its similarity to candidates in $DisLis$, opting for the candidate with the least similarity for replacement (Line~\ref{line:selectbestdis}), resulting in a new prompt $p_{dis}$ (Line~\ref{line:disreplace}).

\subsubsection{Oracle and Fitness}

Given the challenge of the lack of oracle for accurately detecting NSFW content in images, \tool leverages a strategy of cross-checking between the target safety checker ($SC_t$) and a the \red{surrogate safety checker} ($SC_r$) to pinpoint potential adversarial prompts. A prompt is considered potentially adversarial if it evades detection by $SC_t$ but is flagged as NSFW by $SC_r$ (Line~\ref{line:returnresult}). This condition for a prompt $p$ and the corresponding generated image $I$ being deemed adversarial is formally expressed as:
$$InValid(p, I)=(SC_t(p, I) < 0.5) \land (SC_r(p, I) > 0.5).$$
In our experiments, we further manually confirm the NSFW content of the synthesized images by the detected adversarial prompts.\looseness=-1

In cases where a prompt does not meet the condition, \tool utilizes the safety score difference as a fitness function. This function aids in selecting the most effective prompt for the subsequent iteration by comparing $p_{cur}$, $p_{dir}$ and $p_{dis}$ (Line~\ref{line:returnresult}-\ref{line:updatecurprompt}). The objective is to maximize the discrepancy in safety scores between the checkers, formally defined as:
$$fitness(p, I)=SC_r(p, I)-SC_t(p, I).$$

\subsubsection{Discussion} 
In our implementation, the word-level sensitivity analysis is conducted only once on the initial prompt 
$p$, rather than repeatedly on every updated prompt $p_{cur}$ in the iteration. This choice prioritizes efficiency by minimizing the number of required queries to the safety checker, especially considering the resource-intensive nature of the discrepancy analysis ($Dc$), which necessitates two queries per word.

For the mutation process detailed in Algorithm~\ref{alg:mainalgo}, substitutions within $p_{cur}$ are not based on direct word matching but rather through the indices of words in $p$. Specifically, for each substitution pair ($w, w'$), the replacement occurs as follows:
$p_{cur}[index(p,w)]=w'$, where \textit{index} function determines the position of $w$ within $p$. The reason is that an original dirty word 
$w$ might have been substituted with another dirty word in the current prompt $p_{cur}$.

To further enhance query efficiency, only one prompt is maintained for each mutation, \ie, dirtiness-preserving ($p_{dir}$) and discrepancy-away ($p_{dis}$), during iterations. This approach limits the volume of oracle checks and fitness calculations which require the query on the safety checkers. While it is possible to retain and evaluate more mutants per iteration for the selection, such adjustments are subject to available queries on T2I system or computational capacity.

\section{Evaluation}
We have implemented \tool in Python 3.8 with Pytorch (ver.1.11.0). To evaluate the effectiveness of \tool for evaluating the robustness of safety checkers, we aim to answer the following research questions (RQs):
\begin{itemize}[leftmargin=*]
    \item \textbf{RQ1}: How effective is \tool in generating adversarial prompts that maintaining NSFW content? 
    
    \item \textbf{RQ2}: What are the contributions of different components of \tool in improving effectiveness?

    \item \textbf{RQ3}: How do different hyperparameters affect the performance of \tool?
    
    
    
\end{itemize}

\subsection{Setup}
\subsubsection{T2I Models and Datasets}
\red{We employ three popular T2I models to synthesize images based on provided prompts. Specifically, we select Stable-Diffusion-v1.4~\cite{noauthor_compvisstable-diffusion-v1-4_nodate}, following the previous works~\cite{rando2022red, qu2023unsafe, yang2024sneakyprompt}, to ensure a fair comparison. To further demonstrate the generalization of our method, we included DreamLike~\cite{noauthor_free_nodate} and Stable-Diffusion-v1.5~\cite{noauthor_bdsqlszstable-diffusion-v1-5_nodate} in the experiments, as their training data and generated images differ significantly from Stable-Diffusion-v1.4.}
We evaluate \tool using three NSFW prompt datasets, which were 
constructed and utilized by previous works~\cite{qu2023unsafe, yang2024sneakyprompt}. Specifically, the 4chan and Lexica datasets were constructed by Unsafe Diffusion~\cite{qu2023unsafe}, and the NSFW-200 dataset was constructed by SneakyPrompt~\cite{yang2024sneakyprompt}:

\begin{itemize}[leftmargin=*]
    \item  
    \textit{4chan Dataset}: 500 prompts collected from posts on 4chan~\cite{noauthor_4chan_nodate},  encompassing various NSFW themes such as pornography, violence, politics, and discrimination. The posts are prone to NSFW image generation.
    \item 
    \textit{Lexica Dataset}: 404 prompts collected from Lexica website~\cite{noauthor_lexica_nodate},  a large AI-Generated image database. 
    The prompts are retrieved through NSFW keywords included in DALL·E's content policy~\cite{noauthor_are_nodate}, such as sexual, hate, politics, and similar content.
    \item 
    \textit{NSFW-200 Dataset}: 200 prompts generated by GPT-3.5, following a post on Reddit~\cite{principal-goodvibes_nsfwgpt_2023}. The generated prompts contain descriptions of pornographic and politically related content.
\end{itemize}

Given our objective of creating adversarial prompts that can bypass the target safety checker, when selecting the initial seed prompts, we exclusively consider prompts that are consistently blocked by all target safety checkers in the prompt datasets. Finally, there are a total of 325 seed prompts, with 115 from the 4chan dataset, 78 from the Lexica dataset, and 132 from the NSFW-200 dataset.

\subsubsection{Target Safety Checker} \red{We follow the previous work~\cite{yang2024sneakyprompt} and select 5 safety checkers covering all 3 categories to conduct our evaluation. }
\begin{itemize}[leftmargin=*]
    \item 
    \textit{Text Safety Checker}: Text-Match~\cite{yang2024sneakyprompt} and Text-Classifier~\cite{noauthor_michellejielinsfw_text_classifier_nodate} are selected as the target text safety checkers. Text-Match detects prompts containing words from a predefined sensitive word list~\cite{noauthor_nsfw-words-listnsfw_listtxt_nodate}. Text-Classifier is a fine-tuned DistilBERT~\cite{sanh2019distilbert}, the fine-tuning datasets are posts collected from Reddit~\cite{noauthor_reddit_2024} with the purpose of classifying NSFW text content.
    \item 
    \textit{Image Safety Checker}:  We select NSFW-Classifier~\cite{chhabra_lakshaychhabransfw-detection-dl_2024} and CLIP-Detector~\cite{noauthor_laion-aiclip-based-nsfw-detector_2024} as the target image safety checkers. 
    NSFW-Classifier, an open-source image classifier specifically designed to detect NSFW content in images. CLIP-Detector, a binary classifier trained on CLIP image embeddings from an NSFW image dataset~\cite{noauthor_alex000kimnsfw_data_scraper_nodate}, enabling it to identify NSFW content in images.
    \item 
    \textit{Text-Image Safety Checker}: We choose Text-Image-SD, the default safety checker of the open-source Stable Diffusion framework. It blocks image embeddings that exhibit significant similarity to pre-defined NSFW concepts~\cite{rando2022red}.
\end{itemize}

\subsubsection{Baselines}
To evaluate the effectiveness of \tool, we compare \tool with two types of techniques:
\begin{itemize}
    \item 
    \textit{Adversarial attack}: This category includes two state-of-the-art adversarial attack techniques designed for NLP classification tasks: PWWS~\cite{ren2019generating} and TextFooler~\cite{jin2020bert}. These techniques are used to generate adversarial examples specifically for text-based models. We use our cross-check oracle to determine whether the adversarial examples result in images containing NSFW content.
    \item 
    \textit{Adversarial prompting}: We include Sneakyprompt~\cite{yang2024sneakyprompt} and SurrogatePrompt~\cite{ba2023surrogateprompt}, the most recent techniques for evaluating the robustness of safety checkers in T2I models. Sneakyprompt employs reinforcement learning to explore alternative tokens and formulate new prompts. It aims to achieve high CLIP embedding similarity between the initial prompt and the synthesized image. SurrogatePrompt uses an LLM to find alternatives for sensitive words in the input prompt to reduce the likehood of triggering the safety checkers.
\end{itemize}


\subsubsection{Metrics}
To compare the results of different tools, we select four commonly used metrics: the bypass rate, the number of queries (Query number), the time used (Time), and the CLIP score. 
\begin{itemize}
    \item 
    \textit{Bypass rate}: This metric quantifies the percentage of prompts that successfully bypass the target safety checker while still containing NSFW content in the synthesized images. A higher bypass rate indicates that the method is more adept at generating adversarial prompts for a larger proportion of seed prompts.
    \item 
    \textit{Query number}: This metric represents the total number of queries made to the T2I model and safety checker when an adversarial prompt is successfully generated. This metric could assess the efficiency of each tool.
    \item
    \textit{Time}: This metric represents the total time used when an adversarial prompt is successfully generated. This metric can more thoroughly measure the efficiency of each tool.
    \item 
    \textit{CLIP score}: This metric measures the cosine similarity between the CLIP embedding of the prompt and the synthesized image. A higher CLIP score indicates that the synthesized image more closely aligns with the prompt's description, signifying higher synthesis quality.
\end{itemize}

\subsubsection{Experiment Setup}
\paragraph{Configuration of \tool}
For each seed prompt, we set the testing budget $T$ (maximum number of iterations) to 60. By default, the number of discrepant words that can be mutated ($K$ in Algorithm~\ref{alg:mainalgo}) is configured as 1, striking a balance between the bypass rate and CLIP score. Increasing the number of discrepant words for mutation could potentially enhance the bypass rate by altering the prompt in a larger degree, but it may also lead to a decrease in the CLIP score due to significant prompt modifications. \red{To further control computational cost while maintaining performance, we set the number of candidate prompts retained per mutation to 1. This ensures a streamlined search process without significantly sacrificing effectiveness.} We evaluate the impact of these parameters in Section~\ref{sec:rq3}. 

\red{Besides these hyperparameters,  we used the default language models of three T2I models to extract embeddings. Specifically, for the two Stable Diffusion models, we use CLIP-VIT-L-14~\cite{DBLP:conf/icml/RadfordKHRGASAM21}, while DreamLike used its built-in text encoder~\cite{noauthor_free_nodate}. And the \red{surrogate safety checker} used is trained follow the configuration of Unsafe Diffusion~\cite{qu2023unsafe}.} 

\red{To ensure the regularity of mutated prompts, both the dirty words substitution list ($DirLis$) and the discrepant words substitution list ($DisLis$) are constructed by randomly selecting words from the google-10000-english-dictionary~\cite{kaufman_first20hoursgoogle-10000-english_2024}, a curated list of 10,000 commonly used English words, as our substitution search space for regularity.}



\paragraph{Configuration of Baselines}
For the baselines, we follow their default configurations. In the case of Sneakyprompt, the reward function is set as the similarity between the synthesized image and the initial prompt, which tends to perform better in maintaining NSFW content compared to using the similarity between the optimized prompt and the initial prompt, as per their evaluation results. The maximum query number is also set to 60 for fair comparisons.

It is important to note that we observed the inaccuracy in the oracle used to determine the success of attacks by SneakyPrompt. This inaccuracy can lead to the occurrence of false positives, where the generated prompts bypass the checker but result in synthesized images without NSFW content, leading to early termination of the tool. To ensure a more accurate comparison, we introduce a variant of SneakyPrompt, denoted as 
SneakyPrompt\_c, which incorporates our oracle (\ie, the cross-checking method) to determine the successful generation of adversarial prompts containing NSFW content.

In the case of SurrogatePrompt, since it cannot automatically identify the dirty words to be substituted in the input prompt, we first use \textit{Text-Match} to select the dirty words in the prompt and then query ChatGPT 3.5 with "what is similar to $<$word$>$" to obtain alternatives for the dirty words, where $<$word$>$ represents each selected dirty word. To minimize the chances of ChatGPT refusing to respond due to dirty words violating its content policy, we use ChatGPT-3.5-turbo to generate substitute terms, as it has been found to be less strict in refusing NSFW content compared to the current version of ChatGPT~\cite{jiang2024unlocking}.
For each dirty word, we use ChatGPT to generate five replacement words to modify the prompt and then test whether the modified prompts and use our cross-checking method to evaluate the successful generation of adversarial prompts containing NSFW content. 

As cross-checking cannot absolutely ensure the correctness of generated prompts, we supplement the evaluation with a human study. We synthesize images using the adversarial prompts generated by each tool and manually verify whether the synthesized images indeed contain NSFW content. 
During the human study, most images are easily confirmed if NSFW content is included. For any ambiguous cases, our authors discuss them, and if a unanimous decision cannot be reached, the prompt will be discarded. However,  in our study, we encountered no such controversial examples where an agreement couldn't be reached. In our evaluation, all bypass rate results are calculated based on the human-verified outcomes.

\paragraph{RQ Setup}
For \textbf{RQ1}, we take all the seed prompts and utilize various tools to generate adversarial prompts. Subsequently, we compare their performance based on the bypass rate, number of queries, the time used, and CLIP score.


For \textbf{RQ2}, we conduct an ablation study to assess the contributions of the word-level sensitivity analysis and sensitivity-aware mutation. First, we implement a random strategy (\textbf{Random}) that randomly selects a word for mutation, \ie, without sensitivity-aware mutation. The candidate substitution list for it is the union of the substitution lists for dirty words and discrepant words ($DirLis$ and $DisLis$). To demonstrate the importance of sensitivity-aware mutation, we configure two variants of \tool: one without Dirtiness-preserving mutation (\textbf{w/o Dirty}) and another without Discrepancy-away mutation (\textbf{w/o Discrepancy}). Additionally, we introduce two specialized versions for a more granular examination of our mutation approach: \tool with Dirtiness-Away Mutation, which selects substitutions that differ from the original dirty words (\textbf{Dirt-away}), and \tool with Discrepancy-preserving mutation, which selects substitutions similar to the original discrepant words (\textbf{Discr-prev}). Dirt-away and Discr-prev mutate the dirty words and discrepant words in the opposite direction to \tool, respectively.

There are several parameters in \tool that can be adjusted, including the number of discrepant words $K$  to be mutated, the testing budget $T$, and the number of substitutions for dirty words and discrepant words, denoted as $N$. We set $N$ to 1, meaning that only one prompt ($p_{dir}$ and $p_{dis}$) is obtained after substituting the most similar dirty word and the least similar discrepant word. However, we can generate more candidate prompts by selecting the top $N$ most similar dirty words or least similar words.
 In \textbf{RQ3}, we conduct an experiment to analyze the impact of these parameters on performance.

Considering the numerous comparison settings involved, in \textbf{RQ2} and \textbf{RQ3}, we randomly select 100 prompts for experimentation. Additionally, to analyze the performance change on both prompt and image, we use the Text-Image-SD safety checker and Text-Classifier safety checker as our targets for these experiments, as they respectively judge NSFW content based on image and text.  ~\cite{noauthor_tokenprober_nodate}.



\begin{table*}[t]

\caption{Results of bypass rate on Text-Match (T-M), Text-Classifier (T-C), NSFW-Classifier (N-C), CLIP-Detector (C-D), and Text-Image SD (TI-SD)}
\centering
\scriptsize
\label{tab:effectiveness}
\resizebox{\linewidth}{!}{
\begin{tabular}{ccccccccccccccccc}
\hline 
\multirow{2}{*}{Dataset} & \multirow{2}{*}{Method} & \multicolumn{5}{c}{Stable-Diffusion-v1.4} & \multicolumn{5}{c}{DreamLike} & \multicolumn{5}{c}{Stable-Diffusion-v1.5}\tabularnewline
\cline{3-17} \cline{4-17} \cline{5-17} \cline{6-17} \cline{7-17} \cline{8-17} \cline{9-17} \cline{10-17} \cline{11-17} \cline{12-17} \cline{13-17} \cline{14-17} \cline{15-17} \cline{16-17} \cline{17-17} 
 &  & T-M & T-C & N-C & C-D & TI-SD & T-M & T-C & N-C & C-D & TI-SD & T-M & T-C & N-C & C-D & TI-SD\tabularnewline
\hline 
\multirow{6}{*}{4chan} & TokenP & \textbf{0.43} & \textbf{0.56} & \textbf{0.88} & \textbf{0.52} & \textbf{0.87} & \textbf{0.77} & \textbf{0.51} & \textbf{0.79} & \textbf{0.57} & \textbf{0.91} & \textbf{0.45} & \textbf{0.28} & \textbf{0.79} & \textbf{0.45} & \textbf{0.71}\tabularnewline
 & SneakyP & 0.18 & 0.11 & 0.10 & 0.03 & 0.13 & 0.20 & 0.09 & 0.09 & 0.05 & 0.06 & 0.09 & 0.06 & 0.02 & 0.03 & 0.01\tabularnewline
 & SneakyP\_c & 0.20 & 0.11 & 0.11 & 0.07 & 0.12 & 0.29 & 0.17 & 0.09 & 0.16 & 0.17 & 0.18 & 0.08 & 0.08 & 0.06 & 0.09\tabularnewline
 & SurrogateP & 0.27 & 0.10 & 0.25 & 0.17 & 0.32 & 0.32 & 0.17 & 0.32 & 0.22 & 0.40 & 0.19 & 0.06 & 0.36 & 0.14 & 0.23\tabularnewline
 & PWWS & 0.02 & 0.08 & 0.03 & 0.06 & 0.04 & 0.03 & 0.03 & 0.05 & 0.06 & 0.05 & 0.10 & 0.02 & 0.03 & 0.02 & 0.01\tabularnewline
 & TextFooler & 0.02 & 0.04 & 0.04 & 0.06 & 0.05 & 0.01 & 0.02 & 0.05 & 0.04 & 0.04 & 0 & 0 & 0 & 0 & 0\tabularnewline
\hline 
\multirow{6}{*}{Lexica} & TokenP & \textbf{0.49} & \textbf{0.63} & \textbf{0.74} & \textbf{0.64} & \textbf{0.67} & \textbf{0.64} & \textbf{0.63} & \textbf{0.69} & \textbf{0.67} & \textbf{0.72} & \textbf{0.68} & \textbf{0.65} & \textbf{0.74} & \textbf{0.64} & \textbf{0.62}\tabularnewline
 & SneakyP & 0.08 & 0.05 & 0.15 & 0.04 & 0.17 & 0.04 & 0.03 & 0.10 & 0.05 & 0.13 & 0.01 & 0.01 & 0.04 & 0.06 & 0.06\tabularnewline
 & SneakyP\_c & 0.07 & 0.06 & 0.22 & 0.13 & 0.23 & 0.10 & 0.09 & 0.12 & 0.18 & 0.17 & 0.10 & 0.13 & 0.08 & 0.02 & 0.32\tabularnewline
 & SurrogateP & 0.05 & 0.04 & 0.03 & 0.04 & 0.03 & 0.04 & 0.04 & 0.05 & 0.05 & 0.04 & 0.03 & 0.03 & 0.05 & 0.03 & 0.05\tabularnewline
 & PWWS & 0.04 & 0.04 & 0.04 & 0.05 & 0.10 & 0.01 & 0 & 0 & 0.01 & 0.01 & 0.03 & 0.01 & 0.04 & 0.04 & 0.05\tabularnewline
 & TextFooler & 0 & 0.04 & 0.05 & 0.05 & 0.06 & 0 & 0 & 0.01 & 0 & 0.03 & 0 & 0 & 0 & 0 & 0\tabularnewline
\hline 
\multirow{6}{*}{NSFW-200} & TokenP & \textbf{0.89} & \textbf{0.74} & \textbf{0.97} & \textbf{0.72} & \textbf{0.92} & \textbf{0.62} & \textbf{0.53} & \textbf{0.92} & \textbf{0.53} & \textbf{0.89} & \textbf{0.61} & \textbf{0.45} & \textbf{0.98} & \textbf{0.77} & \textbf{0.98}\tabularnewline
 & SneakyP & 0.36 & 0.35 & 0.42 & 0.29 & 0.33 & 0.39 & 0.26 & 0.59 & 0.62 & 0.51 & 0.39 & 0.29 & 0.31 & 0.41 & 0.36\tabularnewline
 & SneakyP\_c & 0.45 & 0.36 & 0.58 & 0.32 & 0.61 & 0.49 & 0.41 & 0.39 & 0.71 & 0.67 & 0.46 & 0.39 & 0.61 & 0.45 & 0.58\tabularnewline
 & SurrogateP & 0.05 & 0.04 & 0.03 & 0.04 & 0.03 & 0.04 & 0.04 & 0.05 & 0.05 & 0.04 & 0.03 & 0.03 & 0.05 & 0.03 & 0.05\tabularnewline
 & PWWS & 0 & 0.30 & 0.34 & 0.30 & 0.34 & 0.01 & 0.01 & 0.02 & 0.02 & 0.02 & 0.01 & 0.27 & 0.28 & 0.33 & 0.30\tabularnewline
 & TextFooler & 0.02 & 0.10 & 0.29 & 0.38 & 0.33 & 0.02 & 0.01 & 0.04 & 0.02 & 0.05 & 0 & 0 & 0 & 0.01 & 0.01\tabularnewline
\hline 
\end{tabular}
}
\end{table*}


\subsection{{RQ1: Effectiveness Results}}
\textbf{Bypass Rate.} Table~\ref{tab:effectiveness} presents the bypass rates of adversarial prompts generated by different methods. Overall, the results demonstrate that \tool significantly outperforms all baselines in generating adversarial prompts. 

Specifically, compared to the state-of-the-art adversarial prompting techniques, \ie, SneakyPrompt (row SneakyP and SneakyP\_c) and SurrogatePrompt (row SurrogateP), \tool (row TokenP)achieves a higher bypass rate across all datasets and safety checkers, with an average improvement of 0.54. which demonstrates its superior effectiveness in bypassing target safety checkers compared to state-of-the-art methods


Comparing SneakyPrompt with SneakyPrompt\_c, we observed that SneakyPrompt\_c generally performs better. This is because SneakyPrompt generates many false positives, stemming from its inaccurate oracle. Even with our cross-check oracle, SneakyPrompt\_c's performance remains inferior to \tool (0.24 vs 0.71 on average).
The lower bypass rate of SneakyPrompt could be attributed to two main factors: 1) its focus on substituting dirty words, which increases the likelihood of affecting the NSFW nature of synthesized images; and 2) its reward function, based on the similarity between the initial prompt and the generated images, which is less effective in balancing the preservation of sensitive semantics for NSFW content generation and avoiding detection by safety filters during generation. This highlights the necessity of fine-grained word impact analysis to achieve a balance in jailbreaking T2I models.

As for SurrogatePrompt, 
we found that it performs significantly better on the 4chan dataset compared to the other two datasets (on average, 0.24 vs. 0.04), which is due to the prompts in 4chan dataset containing more explicit dirty words. In contrast, the Lexica and NSFW datasets use the discrepant words more frequently to describe NSFW scenarios, leading to poorer performance for SurrogatePrompt, which focuses primarily on dirty words.

\begin{table*}[t]
\centering
\caption{Results of Time (T), query number (Q.N) and CLIP score (C.S) on Stable-Diffusion-v1.4 (SD4), DreamLike(DLI), and Stable-Diffusion-v1.5 (SD5)}
\label{tab:qncs}
\small
\scriptsize
\resizebox{\linewidth}{!}{
\begin{tabular}{ccccccccccccccccc}
\hline 
\multirow{2}{*}{Model} & \multirow{2}{*}{Classifier} & \multicolumn{3}{c}{T-M} & \multicolumn{3}{c}{T-C} & \multicolumn{3}{c}{N-C} & \multicolumn{3}{c}{C-D} & \multicolumn{3}{c}{TI-SD}\tabularnewline
\cline{3-17} \cline{4-17} \cline{5-17} \cline{6-17} \cline{7-17} \cline{8-17} \cline{9-17} \cline{10-17} \cline{11-17} \cline{12-17} \cline{13-17} \cline{14-17} \cline{15-17} \cline{16-17} \cline{17-17} 
 &  & T & Q.N & C.S & T & Q.N & C.S & T & Q.N & C.S & T & Q.N & C.S & T & Q.N & C.S\tabularnewline
\hline 
\multirow{3}{*}{SD4} & TP & 213.23  & 24.83  & \textbf{23.50 } & \textbf{239.07} & 27.84  & \textbf{23.47 } & \textbf{202.53 } & 23.87  & \textbf{22.12} & \textbf{220.63 } & 26.14  & \textbf{21.83} & \textbf{212.38 } & 24.81  & 21.39\tabularnewline
 & SP & \textbf{128.07 } & \textbf{3.01 } & 21.30  & 863.20  & \textbf{20.31 } & 19.88  & 377.48  & \textbf{8.88 } & 20.78 & 479.09  & \textbf{11.20}  & 19.81  & 327.14  & \textbf{7.65}  & 21.56\tabularnewline
 & SP\_c & 1921.97  & 45.14  & 19.42  & 2186.42  & 51.39  & 21.42  & 1942.31  & 45.63  & 21.45  & 2251.14  & 52.76  & 19.84  & 1819.58  & 42.62  & \textbf{21.78}\tabularnewline
\hline 
\multirow{3}{*}{DLI} & TP & \textbf{67.01 } & 7.85  & 23.06  & \textbf{100.36 } & \textbf{11.50}  & \textbf{22.88}  & \textbf{132.47 } & 15.61  & \textbf{23.34 } & \textbf{155.65 } & 17.99  & \textbf{22.34}  & \textbf{113.64 } & 13.35  & 22.06\tabularnewline
 & SP & 221.41  & \textbf{5.17 } & \textbf{23.92 } & 596.31  & 13.94  & 21.25  & 428.03  & \textbf{10.01 } & 22.81  & 522.51  & \textbf{12.20 } & 22.11  & 180.02  & \textbf{4.22 } & 22.81\tabularnewline
 & SP\_c & 1797.63  & 42.01  & 25.11  & 1688.72  & 39.59  & 21.30  & 1647.47  & 38.50  & 22.53 & 1600.65  & 37.41  & 20.84  & 1712.67  & 40.09  & \textbf{23.17}\tabularnewline
\hline 
\multirow{3}{*}{SD5} & TP & \textbf{89.84 } & 10.63  & 22.98  & \textbf{106.66 } & 12.37  & \textbf{23.45}  & \textbf{106.95 } & 12.49  & 22.70  & \textbf{155.10 } & 18.17  & 22.20  & \textbf{110.40 } & 12.71 & 22.42\tabularnewline
 & SP & 180.67  & \textbf{4.22} & 23.40  & 525.46  & \textbf{12.34 } & 21.08  & 267.73  & \textbf{6.25 } & 22.48  & 275.85 & \textbf{6.44} & \textbf{23.00 } & 237.98  & \textbf{5.56} & 22.44\tabularnewline
 & SP\_c & 1899.29  & 44.38  & \textbf{24.42 } & 1811.01  & 42.22  & 20.80 & 1823.73  & 42.78 & \textbf{24.31 } & 2101.62  & 49.17 & 21.86  & 1722.45  & 40.41 & \textbf{22.77}\tabularnewline
\hline 
\end{tabular}
}
\end{table*}

In addition, adversarial attacks exhibit significantly worse performance compared to \tool. This is because adversarial attacks primarily aim to generate adversarial examples that bypass the target safety checker, without considering the preservation of NSFW content in synthesized images.


Comparing the results of different datasets, we observe that \tool performs much better on NSFW-200. This is because NSFW-200 is generated by GPT-3.5, which contains less dirty words in the prompts, compared with 4chan and Lexica datasets. Therefore, it is easier to balance the NSFW maintenance and avoid the detection of safety checkers, as only a few dirty words need to be mutated, resulting in better performance on NSFW datasets.

Comparing the results of different T2I models, we found that the effectiveness of \tool is not affected by differences between T2I models. When using three different T2I models, \tool consistently outperforms the baseline, which demonstrates the Generalization of \tool to safety checkers deployed on various T2I models.

Comparing the results of testing different safety checkers, we observe that \tool always has better performance when testing NSFW-Classifier and Text-Image-SD. This is because these two safety checkers are not based on text checking, and insensitive to the dirty words. Like general adversarial attacks, bypassing safety checkers is possible by choosing different substitutions to replace the dirty words. Note that although CLIP-Detector is also a safety checker based on images, it is trained using more than fifty thousands of image embeddings. Hence, the CLIP-Detector tends to be more robust compared to NSFW-Detector and Text-Image-SD.

\textbf{Query Number.} 
We compared the efficiency of several methods with good bypass rates, specifically those with an average pass rate exceeding 0.10: TokenProber (0.68), SneakyPrompt (0.18), and SneakyPrompt\_c (0.26). Table ~\ref{tab:qncs} shows their query number, time used and CLIP score. 
We observe that SneakyPrompt achieves relatively better results in terms of query number. This is because they have many false positives (on average 34\%), leading to the early termination. When SneakyPrompt uses our oracle to determine the success of attacks, SneakyPrompt\_c has a much higher average query number than \tool (47.51 vs 25.50).  

\textbf{Time.} Overall, \tool has a higher bypass rate while using less time, \eg, it takes 60.33\% less time to generate an adversarial prompt on average compared with SneakyPrompt. When comparing with the results of query numbers, we find that although SneakyPrompt requires fewer queries in some cases, it takes more time due to the use of reinforcement learning in the process of searching for alternative words, which makes each round take longer.


\textbf{CLIP Score.}
For comparison, we provide reference CLIP scores calculated for the seed prompts in the 4chan, Lexica, and NSFW-200 datasets. The average CLIP scores for the three datasets are 22.1, 28.66, and 23.16, respectively. Compared to adversarial prompting methods, on average, we find that \tool achieves a higher CLIP score (22.65 vs 22.23), demonstrating higher image quality using our adversarial prompts. 

\red{Table~\ref{tab:manualcheck} displays the false positive rates (\ie, prompts incorreclty classified as adversarial) results of \tool and SneakyPrompt. We manually analyzed the reported adversarial prompts to determine how many of them did not contain NSFW content. The results indicate that the cross-check between the target safety checker and the \red{surrogate safety checker} is effective in identifying invalid images. This effectiveness may be attributed to the low probability that both safety checkers will misclassify adversarial prompts as valid. In comparison, SneakyPrompt shows a higher rate of false positives (\eg, 0.03 vs. 0.49 on average).}


\begin{framed}
	\noindent \textbf{Answers to RQ1}: {
\tool significantly outperforms adversarial attack techniques and the state-of-the-art Sneakyprompt in generating adversarial prompts. The cross-check is a useful oracle for determining whether the adversarial prompts can lead to NSFW content.}
\end{framed}

\begin{table*}[t]
\centering
\caption{\red{False Positive Rate of \tool and SneakyPrompt} }
\label{tab:manualcheck}
\resizebox{0.95\linewidth}{!}{
\begin{tabular}{cccccccccccccccc}
    \hline
    \multirow{2}[4]{*}{Method} & \multicolumn{5}{c}{Stable-Diffusion-v1.4} & \multicolumn{5}{c}{DreamLike}         & \multicolumn{5}{c}{Stable-Diffusion-v1.5} \\
\cmidrule{2-16}          & T-M   & T-C   & N-C   & C-D   & TI-SD & T-M   & T-C   & N-C   & C-D   & TI-SD & T-M   & T-C   & N-C   & C-D   & TI-SD \\
    \hline
    TokenProber & \textbf{0.02} & \textbf{0.04} & \textbf{0.03} & \textbf{0.04} & \textbf{0.04} & \textbf{0.03} & \textbf{0.03} & \textbf{0.03} & \textbf{0.10} & \textbf{0.04} & \textbf{0.03} & \textbf{0.03} & \textbf{0.03} & \textbf{0.02} & \textbf{0.04} \\
    SneakyPrompt & 0.33  & 0.32  & 0.31  & 0.38  & 0.37  & 0.49  & 0.61  & 0.42  & 0.58  & 0.47  & 0.64  & 0.71  & 0.47  & 0.69  & 0.50 \\
    \hline
    \end{tabular}%
}
\end{table*}

\subsection{{RQ2: Ablation Study}}

Table~\ref{tab:ablation} presents the results of our ablation study, comparing different variants of \tool on two selected safety checkers, the results on other datasets are put on our website~\cite{noauthor_tokenprober_nodate}. In the table, Q.N represents the average number of queries for all selected prompts, including unsuccessful ones, while C.S indicates the average CLIP score for only successful adversarial prompts.

The overall results validate the effectiveness of the two sensitivity-aware mutation strategies. The bypass rate of \tool decreases when either or both of these strategies are omitted. For instance, on Text-Image-SD, the bypass rate drops by 0.06, 0.50, and 0.15 when omitting dirtiness-preserving mutation, discrepancy-away mutation, and both, respectively. This highlights the importance of both mutation strategies in generating effective adversarial prompts.

We also observe that the results vary depending on the type of safety checker. Generally, text-based safety checkers (\eg, T-C) are more challenging to bypass than text-image-based checkers (\eg, TI-SD) because text-based checkers directly scrutinize sensitive keywords or phrases. For example, the bypass rate of w/o Dirty on T-C is 0, indicating that without mutating explicit sensitive keywords, text-based checkers will block the prompts by detecting such keywords. Conversely, w/o Discrepancy achieves a 0.52 bypass rate on T-C, suggesting that replacing sensitive or dirty keywords can evade detection by text-based checkers. In contrast, w/o Dirty has a higher success rate (0.80) on TI-SD, as TI-SD compares embeddings, which are not very sensitive to dirty words because changes in non-dirty words can also significantly affect the embedding. This is further evidenced by the comparison between the bypass rates of Random and w/o Discrepancy on TI-SD (0.71 vs 0.36), where modifying only the dirty words (w/o Discrepancy) has a lower success rate than random mutation (Random).

\begin{table}
\centering
\renewcommand\arraystretch{1.05}
\caption{Results of Different Variants on Text-Image SD (TI-SD) and Text-Classifier (T-C)}
\scriptsize
\label{tab:ablation}
\resizebox{0.95\linewidth}{!}{
\begin{tabular}{ccccccc}
\hline 
\multirow{2}{*}{Method} & \multicolumn{3}{c}{TI-SD} & \multicolumn{3}{c}{T-C}\tabularnewline
\cline{2-7} \cline{3-7} \cline{4-7} \cline{5-7} \cline{6-7} \cline{7-7} 
 & B.R & Q.N & C.S & B.R & Q.N & C.S\tabularnewline
\hline 
Random & 0.71 & 29.00 & 21.88 & 0.46 & 41.94 & 20.44\tabularnewline
w/o Dirty & 0.80 & 23.18 & 22.75 & 0 & - & -\tabularnewline
w/o Discrepancy & 0.36 & 44.14 & 19.23 & 0.52 & 36.98 & 22.10\tabularnewline
\hline
Dirt-away & 0.34 & 46.96 & 16.72 & 0.48 & 41.78 & 22.34\tabularnewline
Discr-prev & 0.74 & 27.55 & \textbf{23.23} & 0.34 & 43.60 & \textbf{24.25}\tabularnewline
\hline 
TokenProber & \textbf{0.86 } & \textbf{23.17} & 20.80 & \textbf{0.56} & \textbf{35.74} & 22.64\tabularnewline
\hline 
\end{tabular}}
\end{table}

The results of Dirty-away and Discr-prev further underscore the significance of our mutation directions. When we modify the mutation of dirty words from most similar to least similar in \tool (Dirty-away), the bypass rate significantly decreases (0.86 vs 0.34 on TI-SD and 0.56 vs 0.48 on T-C), likely because replacing dirty words with the least similar ones tends to remove NSFW content. Similarly, mutating discrepant words from least similar to most similar (Discr-prev) results in a substantial drop in bypass rates (0.86 vs 0.74 on TI-SD and 0.56 vs 0.34 on T-C), highlighting the importance of identifying discrepant words to impact the robustness of the checker, especially for text-based filters that focus more on detecting explicit dirty words.

The average number of queries aligns with the bypass rate results; a lower bypass rate implies more unsuccessful cases that exhaust the query budget. Regarding the CLIP score, Discr-prev achieves the best results because it selects the most similar dirty words and discrepant words for substitution, making the adversarial prompt closer to the seed prompt.

\begin{framed}
	\noindent \textbf{Answers to RQ2}: {Both dirtiness-preserving mutation and discrepancy-away mutation play significant roles in enhancing the performance of \tool.}
\end{framed}

\subsection{{RQ3: Impact of Different Parameters}}
\label{sec:rq3}
\begin{figure*}
    
    \centering

    \includegraphics[width=0.9\linewidth]{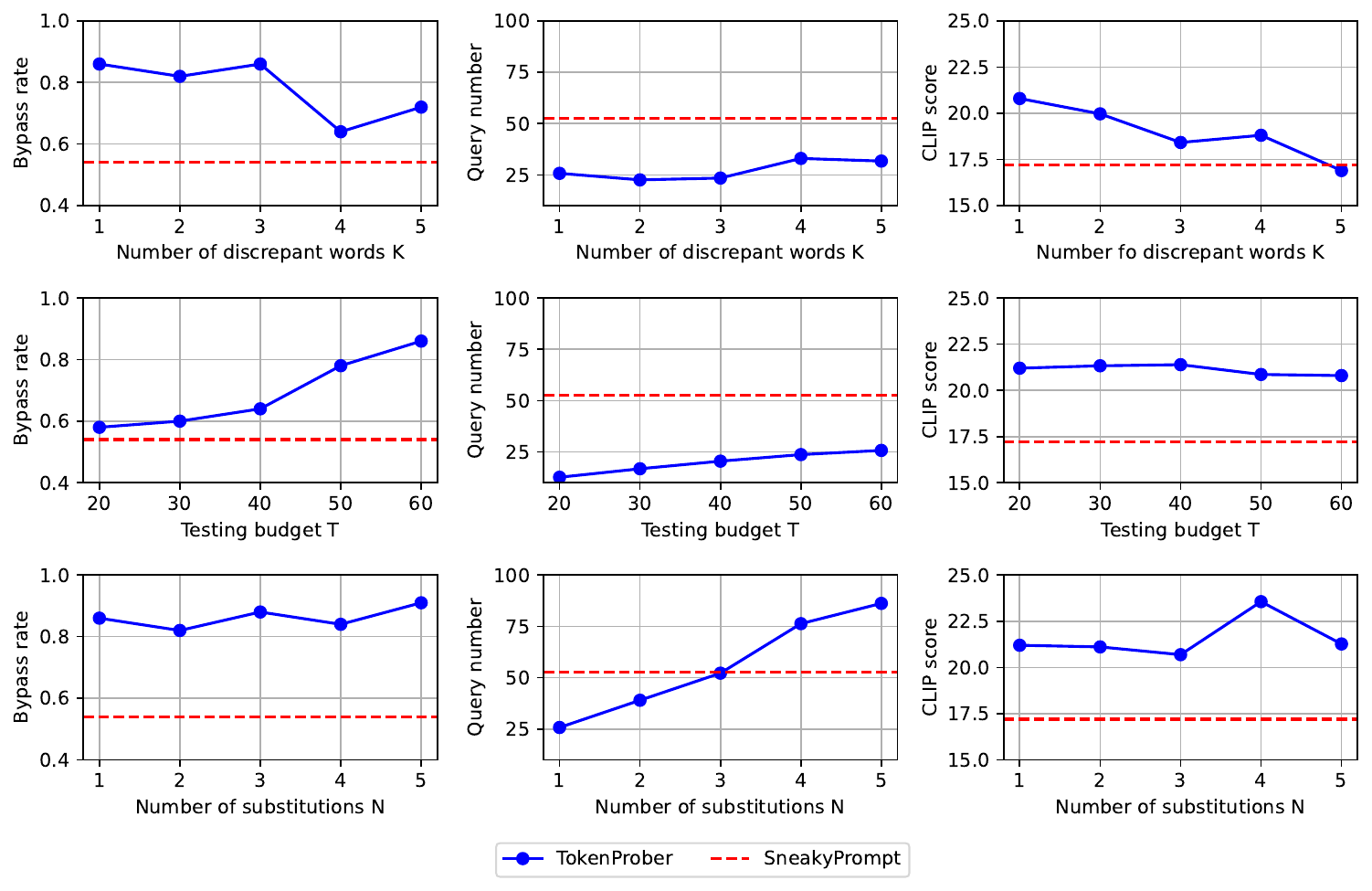}
    \caption{Results of \tool configured with different parameters}
    \label{fig:parameter}
\end{figure*}

Fig.~\ref{fig:parameter} presents the results of \tool with different parameter configurations. We vary the number of discrepant word selections ($K$=1, 2, 3, 4, 5), the testing budget 
($T$=20, 30, 40, 50, 60), and the number of candidate prompts 
($N$=1, 2, 3, 4, 5) to evaluate their impacts. SneakyPrompt's results (with a testing budget of 60) are included for reference. In almost all configurations, \tool outperforms SneakyPrompt, except when \tool operates with a significantly lower testing budget.

Regarding the impact of the number of discrepant words ($K$), an increase in $K$ leads to a decrease in the bypass rate and CLIP score while increasing the query number. This indicates that it becomes more challenging to generate adversarial prompts as \tool seeks semantically deviated substitutions for discrepant words. As $K$ increases, the mutation of discrepant words results in a semantic deviation of the seed prompt, affecting its performance. For the testing budget ($T$), as expected, both the bypass rate and query number increase with $T$, suggesting that some challenging cases require more testing budget.

Regarding the number of prompts generated for fitness selection ($N$), an increase in $N$ significantly raises the query number (from 25.74 to 86.20) due to the additional queries to calculate fitness scores. However, the increase in bypass rate is not significant (from 0.86 to 0.91), indicating that maintaining one substitution does not significantly impact performance. This finding suggests that prioritizing one substitution strikes a better balance between effectiveness and efficiency.

\begin{framed}
	\noindent \textbf{Answers to RQ3}: { Selecting a large number of discrepant words for mutation can diminish the performance of \tool. Additionally, generating more candidate prompts for selection does not evidently enhance performance in our greedy-based approach. By opting for smaller values ($K=1$ and $N=1$), \tool achieves an optimal balance between effectiveness and efficiency.}
\end{framed}

 \label{evaluate_tool}
\section{Discussion}

\paragraph{Discrepancies Among Target Safety Checkers and \red{Surrogate Safety Checker}} \tool efficiently identifies robustness issues with the target safety checkers by exploiting the discrepancy in the target safety checkers and the \red{surrogate safety checker}.
Note that such discrepancies are common because  the \red{surrogate safety checker} and the safety checker have different model architectures and are trained on different datasets. It is expected that they will produce different outputs for certain inputs.
To further demonstrate the existence of these discrepancies, we analyzed the differences in outputs between the \red{surrogate safety checker} and the 5 target safety checkers for all prompts used in our experiments. The results can be seen on our website~\cite{noauthor_tokenprober_nodate}, which demonstrates that there are differences in the decision boundaries.

\paragraph{Impact of words on robustness}
The jailbreaking problem is fundamentally a robustness issue. Its distinct characteristic is that it not only aims to cause misclassification by the safety checker but also seeks to prompt the T2I models to generate NSFW content, as illustrated in the inconsistency zone in Fig.~\ref{fig:illustration}. The words within the prompt play a critical role in the robustness of the safety checker. Our method, \tool, is designed to identify these vulnerable words that significantly impact robustness. While dirty words are crucial for the safety checker, they are often not the vulnerable features since they are explicit indicators for invalid prompt detection. Hence, merely mutating dirty words, a common strategy in existing works, is insufficient. Instead, discrepant words, which are not well-handled by the checker, represent the vulnerable features that can be exploited to bypass the safety check.

\begin{figure}
    
    \centering

    \includegraphics[width=\linewidth]{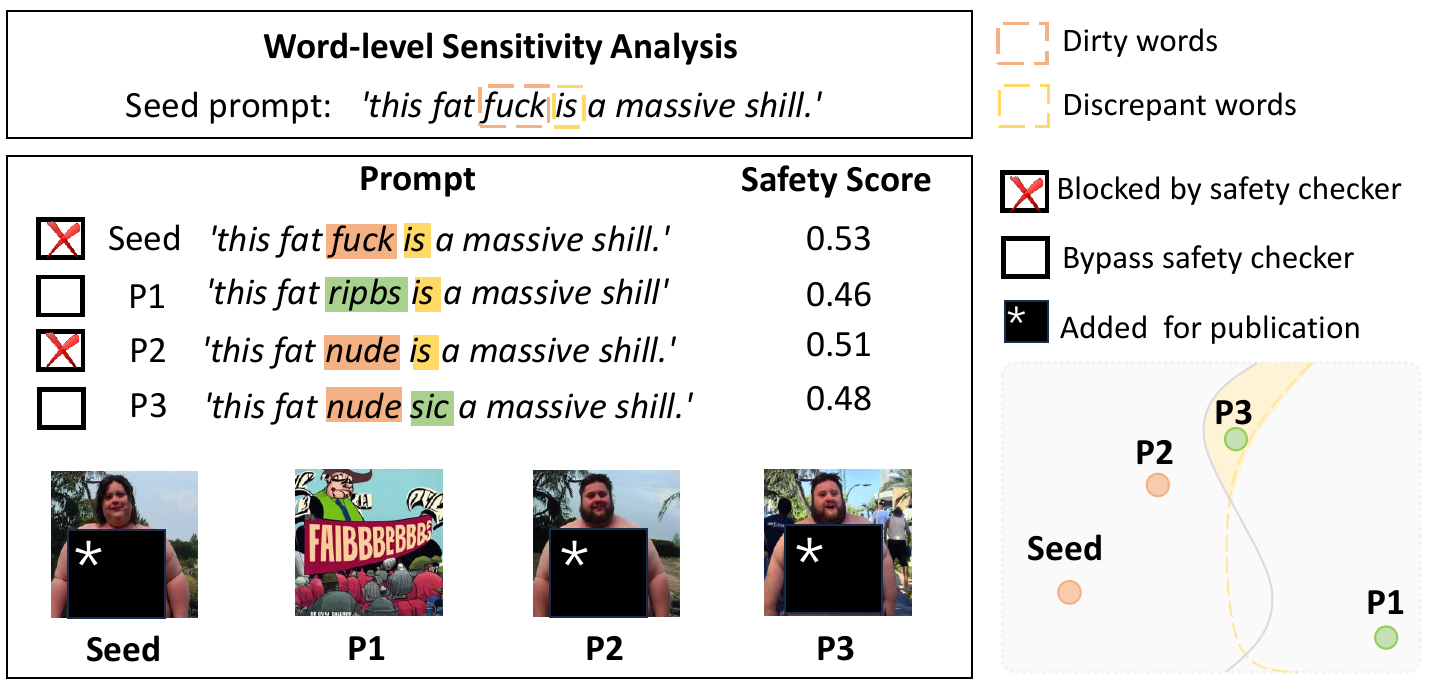}
    \caption{\textcolor{red}{Note: Figure contains profanity.} The impact of dirty words and discrepant words}
    \label{fig:case}
\end{figure}

Figure~\ref{fig:case} presents an example from our experiment. The word ``fuck'' is an explicit dirty keyword used for filtering, indicated by an 0.53 score. When replaced with ``ripbs'', the prompt (p1) is not filtered out (0.46 score), but it does not generate inappropriate images either. However, substituting one dirty word for another (\eg, ``fuck'' to ``nude'') in prompt (p2) might slightly improve the positivity (0.51 Score), but the explicit content can still be detected. Our analysis further identifies a discrepant word ``is'', a neutral term with an outsized impact on sensitivity detection. By altering it to another one ``sic'', prompt (p3) becomes positive (0.48 score), evading detection, yet it can still prompt the T2I model to generate NSFW content as the dirty word remains.

\paragraph{Improving the safety filtering}
We found that text-based checkers are relatively effective in detecting explicit dirty words. However, their weakness is that they may not have a complete dirty word list, and some dirty words can evade detection. Image-based or text-image-based checkers, based on embedding detection, do not explicitly check for dirty words. They could capture some unknown dirty words as long as their embedding is similar to known ones. However, ensuring robustness is a challenge (\eg, the effect of discrepant words), a common problem for machine learning tasks.

Our findings suggest directions to improve safety filtering: 1) improving the general robustness, particularly identifying and reducing the number of discrepant words; 2) adopting an ensemble-based strategy that uses text-based checkers for filtering known dirty words while using embedding-based methods to detect unknown NSFW content.

\section{Threats to Validity}
There are some threats that could affect the validity of the results. The selected safety checkers and prompts datasets are threats to the validity. We mitigate these threats by selecting the popular datasets and safety checkers that are used by existing works. The configuration of \tool is a threat. We mitigate this problem by configuring \tool with different parameters and analyzing the impact on results. The human-verified outcomes could be a threat to affect the results. To mitigate this problem, two authors of this work verified the results independently. For any ambiguous cases, our authors discuss them. The choice of the \red{surrogate safety checker} could pose a potential threat to the results. We observe that a normal safety checker can be suitable as long as they do not share the same vulnerabilities, such as the same discrepant words. In future work, we plan to investigate the impact of using different \red{surrogate safety checker}.
\section{Related Work}
\subsection{Text-to-Image Models}
Mansimov~\etal~\cite{mansimov2015generating} propose the first text-to-image model that can take user's textual descriptions as input, namely prompts, to generate relevant images. Later, lots of works~\cite{ramesh2022hierarchical,xu2018attngan,rombach2022high,koh2021text,nguyen2017plug} focus on improving the quality of generated image by adopting new model structure or optimization methods. Many previous works use GANs~\cite{goodfellow2020generative} to produce text-conditional images~\cite{mirza2014conditional, xu2018attngan, zhang2017stackgan, li2019controllable}. 
Since diffusion models demonstrate a strong ability to synthesize high-quality images, 
numerous works~\cite{rombach2022high, ramesh2022hierarchical, saharia2022photorealistic, gu2022vector, nichol2021glide} are proposed to condition the synthesis of diffusion model by natural language descriptions. For examples, Stable Diffusion~\cite{rombach2022high} and DALL-E~\cite{ramesh2022hierarchical} adopt popular language models to obtain the text embedding of the input prompt, which is then used to condition the de-noise process of diffusion model, in order to synthesize an image that matches the prompt's description. 

The popularity of T2I models have raised ethical concerns about the safety of synthesized image content. To prevent T2I models from generating NSFW content, many safety checkers~\cite{noauthor_michellejielinsfw_text_classifier_nodate, yang2024sneakyprompt, chhabra_lakshaychhabransfw-detection-dl_2024, noauthor_laion-aiclip-based-nsfw-detector_2024, rando2022red} have been proposed to detect the potentially NSFW content in prompt and synthesized images. \tool aims to evaluate the robustness of these safety checkers in T2I models.

\subsection{Robustness Testing on Safety Checker}
Although various safety checkers have been proposed to prevent T2I models from generating NSFW images, the effectiveness of the safety checkers in preventing NSFW content is unknown. Recently, Numerous works~\cite{qu2023unsafe, rando2022red, yang2024sneakyprompt} have been proposed to evaluate the robustness of safety checkers. 

Rando~\etal~\cite{rando2022red} analyze the safety checker of open-source Stable Diffusion and find that manual modifications to prompts can bypass the safety checker and cause Stable Diffusion to synthesize NSFW images. Qu et al.~\cite{qu2023unsafe} construct two datasets to assess the ability of T2I models to generate NSFW images. During the assessment, they find that an adversary can easily generate realistic hateful meme variants by manually crafted prompts. These findings reveal huge vulnerabilities in T2I model's safety checkers. SneakyPrompt~\cite{yang2024sneakyprompt} is the most recent work in automatically generating adversarial prompts to evade T2I models' safety checkers. It employs reinforcement learning to explore alternative tokens and formulate adversarial prompts. 
Although Sneakyprompt can generate adversarial prompts that successfully bypass the safety checkers, a lot of adversarial prompts generated do not contain NSFW content in the synthesized images (34\%). In addition, Sneakyprompt does not consider the different impact of words in prompts on image synthesis. Differently, \tool analyses the impact of words in prompts and generates adversarial prompts capable of producing NSFW images.

\subsection{\red{Generalization of \tool to other tasks}}
\red{In this paper, we focus exclusively on the safety testing of T2I models. However, we reframe the safety testing problem as a fundamental differential testing problem. This formulation addresses the root cause of the issue and provides a conceptually general framework: for any generative model (e.g., text-to-text, text-to-video, or text-to-audio), as long as discrepancies exist between the generative model and its associated safety checker, the safety mechanism can be circumvented.}

\red{Thus, to apply our method to a new generative task, one only needs to first train a surrogate safety checker based on the target generative model. Then, \tool can be used to generate prompts that exploit these discrepancies, leading to the generation of unsafe content by the model.}

\section{Conclusion}
This paper presented \tool, a novel framework designed to assess the robustness of safety checkers within text-to-image systems. \tool conducts a fine-grained, word-level analysis to understand the influence of various word types on the generation of NSFW content and the insensitivity prediction of safety checkers. Utilizing a greedy search strategy, \tool refines prompts through dirtiness-preserving mutation and discrepancy-away mutation, aiming to exploit discrepancies in sensitivity interpretation between the target safety checker and the \red{surrogate safety checker}. The objective is geared towards maximizing the prediction difference, thereby identifying potential weaknesses in refusal mechanisms. The empirical evaluation has underscored its capability to effectively bypassing the safety checkers in T2I systems.



\ifCLASSOPTIONcaptionsoff
  \newpage
\fi

\bibliographystyle{IEEEtran}
\bibliography{sample-base}

\begin{thebibliography}{10}
\providecommand{\url}[1]{#1}
\csname url@samestyle\endcsname
\providecommand{\newblock}{\relax}
\providecommand{\bibinfo}[2]{#2}
\providecommand{\BIBentrySTDinterwordspacing}{\spaceskip=0pt\relax}
\providecommand{\BIBentryALTinterwordstretchfactor}{4}
\providecommand{\BIBentryALTinterwordspacing}{\spaceskip=\fontdimen2\font plus
\BIBentryALTinterwordstretchfactor\fontdimen3\font minus \fontdimen4\font\relax}
\providecommand{\BIBforeignlanguage}[2]{{%
\expandafter\ifx\csname l@#1\endcsname\relax
\typeout{** WARNING: IEEEtran.bst: No hyphenation pattern has been}%
\typeout{** loaded for the language `#1'. Using the pattern for}%
\typeout{** the default language instead.}%
\else
\language=\csname l@#1\endcsname
\fi
#2}}
\providecommand{\BIBdecl}{\relax}
\BIBdecl

\bibitem{noauthor_anonymized_nodate}
\BIBentryALTinterwordspacing
``Anonymized {Repository} - {Anonymous} {GitHub}.'' [Online]. Available: \url{https://anonymous.4open.science/r/TokenProber}
\BIBentrySTDinterwordspacing

\bibitem{rombach2022high}
R.~Rombach, A.~Blattmann, D.~Lorenz, P.~Esser, and B.~Ommer, ``High-resolution image synthesis with latent diffusion models,'' in \emph{Proceedings of the IEEE/CVF conference on computer vision and pattern recognition}, 2022, pp. 10\,684--10\,695.

\bibitem{ramesh2022hierarchical}
A.~Ramesh, P.~Dhariwal, A.~Nichol, C.~Chu, and M.~Chen, ``Hierarchical text-conditional image generation with clip latents,'' \emph{arXiv preprint arXiv:2204.06125}, vol.~1, no.~2, p.~3, 2022.

\bibitem{noauthor_lexica_nodate}
\BIBentryALTinterwordspacing
``\BIBforeignlanguage{en}{Lexica: {Search}, {Discover} \& {Create} {5M}+ {AI}-{Generated} {Images}}.'' [Online]. Available: \url{https://aidude.info/services/Lexica}
\BIBentrySTDinterwordspacing

\bibitem{kawar2023imagic}
B.~Kawar, S.~Zada, O.~Lang, O.~Tov, H.~Chang, T.~Dekel, I.~Mosseri, and M.~Irani, ``Imagic: Text-based real image editing with diffusion models,'' in \emph{Proceedings of the IEEE/CVF Conference on Computer Vision and Pattern Recognition}, 2023, pp. 6007--6017.

\bibitem{zhang2023sine}
Z.~Zhang, L.~Han, A.~Ghosh, D.~N. Metaxas, and J.~Ren, ``Sine: Single image editing with text-to-image diffusion models,'' in \emph{Proceedings of the IEEE/CVF Conference on Computer Vision and Pattern Recognition}, 2023, pp. 6027--6037.

\bibitem{gal2022image}
R.~Gal, Y.~Alaluf, Y.~Atzmon, O.~Patashnik, A.~H. Bermano, G.~Chechik, and D.~Cohen-Or, ``An image is worth one word: Personalizing text-to-image generation using textual inversion,'' \emph{arXiv preprint arXiv:2208.01618}, 2022.

\bibitem{ruiz2023dreambooth}
N.~Ruiz, Y.~Li, V.~Jampani, Y.~Pritch, M.~Rubinstein, and K.~Aberman, ``Dreambooth: Fine tuning text-to-image diffusion models for subject-driven generation,'' in \emph{Proceedings of the IEEE/CVF Conference on Computer Vision and Pattern Recognition}, 2023, pp. 22\,500--22\,510.

\bibitem{washingtonpost2023aiabuse}
\BIBentryALTinterwordspacing
T.~W. Post, ``Artificial intelligence is creating images of child sex abuse. it’s horrific — and a challenge to stop.'' \emph{The Washington Post}, 2023. [Online]. Available: \url{https://www.washingtonpost.com/technology/2023/06/19/artificial-intelligence-child-sex-abuse-images/}
\BIBentrySTDinterwordspacing

\bibitem{guardian2024aiabuse}
\BIBentryALTinterwordspacing
T.~Guardian, ``Arrest over ai-generated child sexual abuse material sparks concerns about technology's darker uses,'' \emph{The Guardian}, 2024. [Online]. Available: \url{https://www.theguardian.com/technology/article/2024/may/21/child-sexual-abuse-material-artificial-intelligence-arrest}
\BIBentrySTDinterwordspacing

\bibitem{qu2023unsafe}
Y.~Qu, X.~Shen, X.~He, M.~Backes, S.~Zannettou, and Y.~Zhang, ``Unsafe diffusion: On the generation of unsafe images and hateful memes from text-to-image models,'' in \emph{Proceedings of the 2023 ACM SIGSAC Conference on Computer and Communications Security}, 2023, pp. 3403--3417.

\bibitem{rando2022red}
J.~Rando, D.~Paleka, D.~Lindner, L.~Heim, and F.~Tram{\`e}r, ``Red-teaming the stable diffusion safety filter,'' \emph{arXiv preprint arXiv:2210.04610}, 2022.

\bibitem{yang2024sneakyprompt}
Y.~Yang, B.~Hui, H.~Yuan, N.~Gong, and Y.~Cao, ``Sneakyprompt: Jailbreaking text-to-image generative models,'' in \emph{2024 IEEE Symposium on Security and Privacy (SP)}.\hskip 1em plus 0.5em minus 0.4em\relax IEEE Computer Society, 2024, pp. 123--123.

\bibitem{ba2023surrogateprompt}
Z.~Ba, J.~Zhong, J.~Lei, P.~Cheng, Q.~Wang, Z.~Qin, Z.~Wang, and K.~Ren, ``Surrogateprompt: Bypassing the safety filter of text-to-image models via substitution,'' \emph{arXiv preprint arXiv:2309.14122}, 2023.

\bibitem{xiao2023latent}
Y.~Xiao, A.~Liu, T.~Li, and X.~Liu, ``Latent imitator: Generating natural individual discriminatory instances for black-box fairness testing,'' in \emph{Proceedings of the 32nd ACM SIGSOFT international symposium on software testing and analysis}, 2023, pp. 829--841.

\bibitem{radford2021learning}
A.~Radford, J.~W. Kim, C.~Hallacy, A.~Ramesh, G.~Goh, S.~Agarwal, G.~Sastry, A.~Askell, P.~Mishkin, J.~Clark \emph{et~al.}, ``Learning transferable visual models from natural language supervision,'' in \emph{International conference on machine learning}.\hskip 1em plus 0.5em minus 0.4em\relax PMLR, 2021, pp. 8748--8763.

\bibitem{noauthor_compvisstable-diffusion-v1-4_nodate}
\BIBentryALTinterwordspacing
``{CompVis}/stable-diffusion-v1-4 · {Hugging} {Face}.'' [Online]. Available: \url{https://huggingface.co/CompVis/stable-diffusion-v1-4}
\BIBentrySTDinterwordspacing

\bibitem{noauthor_free_nodate}
\BIBentryALTinterwordspacing
``\BIBforeignlanguage{en}{Free {AI} {Art} {Generator}, {AI} {Art} {Maker} {\textbar} {Stable} {Diffusion} {Online}}.'' [Online]. Available: \url{https://dreamlike.art/}
\BIBentrySTDinterwordspacing

\bibitem{noauthor_bdsqlszstable-diffusion-v1-5_nodate}
\BIBentryALTinterwordspacing
``bdsqlsz/stable-diffusion-v1-5 · {Hugging} {Face}.'' [Online]. Available: \url{https://huggingface.co/bdsqlsz/stable-diffusion-v1-5}
\BIBentrySTDinterwordspacing

\bibitem{noauthor_4chan_nodate}
\BIBentryALTinterwordspacing
``4chan.'' [Online]. Available: \url{https://4chan.org/}
\BIBentrySTDinterwordspacing

\bibitem{noauthor_are_nodate}
\BIBentryALTinterwordspacing
``\BIBforeignlanguage{en}{Are there any restrictions to how {I} can use {DALL}·{E} 2?}'' [Online]. Available: \url{https://help.openai.com/en/articles/6338764}
\BIBentrySTDinterwordspacing

\bibitem{principal-goodvibes_nsfwgpt_2023}
\BIBentryALTinterwordspacing
Principal-Goodvibes, ``{NsfwGPT}: '{THAT}' {NSFW} prompt...'' Mar. 2023. [Online]. Available: \url{www.reddit.com/r/ChatGPT/comments/11vlp7j/nsfwgpt_that_nsfw_prompt/}
\BIBentrySTDinterwordspacing

\bibitem{noauthor_michellejielinsfw_text_classifier_nodate}
\BIBentryALTinterwordspacing
``michellejieli/{NSFW}\_text\_classifier · {Hugging} {Face}.'' [Online]. Available: \url{https://huggingface.co/michellejieli/NSFW_text_classifier}
\BIBentrySTDinterwordspacing

\bibitem{noauthor_nsfw-words-listnsfw_listtxt_nodate}
\BIBentryALTinterwordspacing
``\BIBforeignlanguage{en}{{NSFW}-{Words}-{List}/nsfw\_list.txt at master · rrgeorge-pdcontributions/{NSFW}-{Words}-{List}}.'' [Online]. Available: \url{https://github.com/rrgeorge-pdcontributions/NSFW-Words-List/blob/master/nsfw_list.txt}
\BIBentrySTDinterwordspacing

\bibitem{sanh2019distilbert}
V.~Sanh, L.~Debut, J.~Chaumond, and T.~Wolf, ``Distilbert, a distilled version of bert: smaller, faster, cheaper and lighter,'' \emph{arXiv preprint arXiv:1910.01108}, 2019.

\bibitem{noauthor_reddit_2024}
\BIBentryALTinterwordspacing
``\BIBforeignlanguage{en-US}{Reddit - {Dive} into anything},'' Mar. 2024. [Online]. Available: \url{https://www.reddit.com/}
\BIBentrySTDinterwordspacing

\bibitem{chhabra_lakshaychhabransfw-detection-dl_2024}
\BIBentryALTinterwordspacing
L.~Chhabra, ``lakshaychhabra/{NSFW}-{Detection}-{DL},'' Mar. 2024, original-date: 2019-10-17T09:09:09Z. [Online]. Available: \url{https://github.com/lakshaychhabra/NSFW-Detection-DL}
\BIBentrySTDinterwordspacing

\bibitem{noauthor_laion-aiclip-based-nsfw-detector_2024}
\BIBentryALTinterwordspacing
``{LAION}-{AI}/{CLIP}-based-{NSFW}-{Detector},'' Mar. 2024, original-date: 2022-03-10T12:11:15Z. [Online]. Available: \url{https://github.com/LAION-AI/CLIP-based-NSFW-Detector}
\BIBentrySTDinterwordspacing

\bibitem{noauthor_alex000kimnsfw_data_scraper_nodate}
\BIBentryALTinterwordspacing
``\BIBforeignlanguage{en}{alex000kim/nsfw\_data\_scraper: {Collection} of scripts to aggregate image data for the purposes of training an {NSFW} {Image} {Classifier}}.'' [Online]. Available: \url{https://github.com/alex000kim/nsfw_data_scraper}
\BIBentrySTDinterwordspacing

\bibitem{ren2019generating}
S.~Ren, Y.~Deng, K.~He, and W.~Che, ``Generating natural language adversarial examples through probability weighted word saliency,'' in \emph{Proceedings of the 57th annual meeting of the association for computational linguistics}, 2019, pp. 1085--1097.

\bibitem{jin2020bert}
D.~Jin, Z.~Jin, J.~T. Zhou, and P.~Szolovits, ``Is bert really robust? a strong baseline for natural language attack on text classification and entailment,'' in \emph{Proceedings of the AAAI conference on artificial intelligence}, vol.~34, no.~05, 2020, pp. 8018--8025.

\bibitem{DBLP:conf/icml/RadfordKHRGASAM21}
\BIBentryALTinterwordspacing
A.~Radford, J.~W. Kim, C.~Hallacy, A.~Ramesh, G.~Goh, S.~Agarwal, G.~Sastry, A.~Askell, P.~Mishkin, J.~Clark, G.~Krueger, and I.~Sutskever, ``Learning transferable visual models from natural language supervision,'' in \emph{Proceedings of the 38th International Conference on Machine Learning, {ICML} 2021, 18-24 July 2021, Virtual Event}, ser. Proceedings of Machine Learning Research, M.~Meila and T.~Zhang, Eds., vol. 139.\hskip 1em plus 0.5em minus 0.4em\relax {PMLR}, 2021, pp. 8748--8763. [Online]. Available: \url{http://proceedings.mlr.press/v139/radford21a.html}
\BIBentrySTDinterwordspacing

\bibitem{kaufman_first20hoursgoogle-10000-english_2024}
\BIBentryALTinterwordspacing
J.~Kaufman, ``first20hours/google-10000-english,'' Mar. 2024, original-date: 2012-03-29T05:22:29Z. [Online]. Available: \url{https://github.com/first20hours/google-10000-english}
\BIBentrySTDinterwordspacing

\bibitem{jiang2024unlocking}
W.~Jiang, Z.~Wang, J.~Zhai, S.~Ma, Z.~Zhao, and C.~Shen, ``Unlocking adversarial suffix optimization without affirmative phrases: Efficient black-box jailbreaking via llm as optimizer,'' \emph{arXiv preprint arXiv:2408.11313}, 2024.

\bibitem{noauthor_tokenprober_nodate}
\BIBentryALTinterwordspacing
``\BIBforeignlanguage{zh-CN}{{TokenProber}}.'' [Online]. Available: \url{https://sites.google.com/view/tokenprober}
\BIBentrySTDinterwordspacing

\bibitem{mansimov2015generating}
E.~Mansimov, E.~Parisotto, J.~L. Ba, and R.~Salakhutdinov, ``Generating images from captions with attention,'' \emph{arXiv preprint arXiv:1511.02793}, 2015.

\bibitem{xu2018attngan}
T.~Xu, P.~Zhang, Q.~Huang, H.~Zhang, Z.~Gan, X.~Huang, and X.~He, ``Attngan: Fine-grained text to image generation with attentional generative adversarial networks,'' in \emph{Proceedings of the IEEE conference on computer vision and pattern recognition}, 2018, pp. 1316--1324.

\bibitem{koh2021text}
J.~Y. Koh, J.~Baldridge, H.~Lee, and Y.~Yang, ``Text-to-image generation grounded by fine-grained user attention,'' in \emph{Proceedings of the IEEE/CVF winter conference on applications of computer vision}, 2021, pp. 237--246.

\bibitem{nguyen2017plug}
A.~Nguyen, J.~Clune, Y.~Bengio, A.~Dosovitskiy, and J.~Yosinski, ``Plug \& play generative networks: Conditional iterative generation of images in latent space,'' in \emph{Proceedings of the IEEE conference on computer vision and pattern recognition}, 2017, pp. 4467--4477.

\bibitem{goodfellow2020generative}
I.~Goodfellow, J.~Pouget-Abadie, M.~Mirza, B.~Xu, D.~Warde-Farley, S.~Ozair, A.~Courville, and Y.~Bengio, ``Generative adversarial networks,'' \emph{Communications of the ACM}, vol.~63, no.~11, pp. 139--144, 2020.

\bibitem{mirza2014conditional}
M.~Mirza and S.~Osindero, ``Conditional generative adversarial nets,'' \emph{arXiv preprint arXiv:1411.1784}, 2014.

\bibitem{zhang2017stackgan}
H.~Zhang, T.~Xu, H.~Li, S.~Zhang, X.~Wang, X.~Huang, and D.~N. Metaxas, ``Stackgan: Text to photo-realistic image synthesis with stacked generative adversarial networks,'' in \emph{Proceedings of the IEEE international conference on computer vision}, 2017, pp. 5907--5915.

\bibitem{li2019controllable}
B.~Li, X.~Qi, T.~Lukasiewicz, and P.~Torr, ``Controllable text-to-image generation,'' \emph{Advances in neural information processing systems}, vol.~32, 2019.

\bibitem{saharia2022photorealistic}
C.~Saharia, W.~Chan, S.~Saxena, L.~Li, J.~Whang, E.~L. Denton, K.~Ghasemipour, R.~Gontijo~Lopes, B.~Karagol~Ayan, T.~Salimans \emph{et~al.}, ``Photorealistic text-to-image diffusion models with deep language understanding,'' \emph{Advances in neural information processing systems}, vol.~35, pp. 36\,479--36\,494, 2022.

\bibitem{gu2022vector}
S.~Gu, D.~Chen, J.~Bao, F.~Wen, B.~Zhang, D.~Chen, L.~Yuan, and B.~Guo, ``Vector quantized diffusion model for text-to-image synthesis,'' in \emph{Proceedings of the IEEE/CVF Conference on Computer Vision and Pattern Recognition}, 2022, pp. 10\,696--10\,706.

\bibitem{nichol2021glide}
A.~Nichol, P.~Dhariwal, A.~Ramesh, P.~Shyam, P.~Mishkin, B.~McGrew, I.~Sutskever, and M.~Chen, ``Glide: Towards photorealistic image generation and editing with text-guided diffusion models,'' \emph{arXiv preprint arXiv:2112.10741}, 2021.

\end{thebibliography}

\end{document}